\documentclass[bibnotes,twocolumn]{revtex4}

\usepackage[dvips]{graphicx}
\usepackage{amssymb}
\usepackage{xspace}

\newcommand{\ybco}{\ensuremath{\mathrm{YBa_2Cu_3O_{7-\delta}}}\xspace}

\begin{document}

\title{Transport and Noise Characteristics of Submicron High-Temperature
  Superconductor Grain-Boundary Junctions}

\author{F.~Herbstritt}
\author{T.~Kemen}
\affiliation{II.~Physikalisches Institut, Universit{\"a}t zu K{\"o}ln,
Z{\"u}lpicherstr.~77, D - 50937 K{\"o}ln, Germany}
\author{L.~Alff}
\author{A.~Marx\setcounter{footnote}{0}}
\email{Achim.Marx@wmi.badw.de}
\author{R.~Gross}
\affiliation{II.~Physikalisches Institut, Universit{\"a}t zu K{\"o}ln,
Z{\"u}lpicherstr.~77, D - 50937 K{\"o}ln, Germany}
\affiliation{Walther-Mei{\ss}ner-Institut f{\"u}r Tieftemperaturforschung,
Walther-Mei{\ss}ner Str. 8, D - 85748 Garching, Germany}

\begin{abstract}
  We have investigated the transport and noise properties of submicron \ybco
  bicrystal grain-boundary junctions prepared using electron beam lithography.
  The junctions show an increased conductance for low voltages reminiscent of
  Josephson junctions having a barrier with high transmissivity. The voltage
  noise spectra are dominated by a few Lorentzian components. At low
  temperatures clear two-level random telegraph switching (RTS) signals are
  observable in the voltage vs time traces. We have investigated the
  temperature and voltage dependence of individual fluctuators both from
  statistical analysis of voltage vs time traces and from fits to noise
  spectra. A transition from tunneling to thermally activated behavior of
  individual fluctuators was clearly observed. The experimental results
  support the model of charge carrier traps in the barrier region.
\end{abstract}

\date{\today}

\maketitle

Grain boundary junctions (GBJs) in high temperature superconductors (HTS) are
widely used for the realization of Josephson junctions\cite{Gross97,Koelle99}.
Despite their simple fabrication there is up to now no consensus concerning
the transport properties of GBJs\cite{Alff98a}. The investigation of low
frequency noise has turned out to be a valuable tool in clarifying the charge
transport
mechanism\cite{Kawasaki92,Kemen99,Marx99,Marx97,Marx97a,Marx95,Marx95a,Hao96a}.
Numerous studies proved that the large amount of $1/f$-noise is due to a high
density of charge trapping states in the grain boundary barrier.
Unfortunately, there is up to now no detailed model for the microscopic nature
of these traps.  The analysis of the dynamics of individual charge traps
should provide valuable information to overcome this
drawback\cite{Ralls91,Rogers85}. Due to the high density of traps in HTS
Josephson junctions the fabrication of small area junctions is required to
achieve a situation where only a single fluctuator is dominating the junction
dynamics.  Therefore, we have fabricated GBJs with a junction area down to
$0.01\,\mu$m$^2$ using electron beam lithography. These junctions allowed for
the investigation of individual fluctuators in a wide range of temperatures
and voltage.

\ybco thin films with a thickness of 20 to 30\,nm were deposited on SrTiO$_3$
$[001]$ bicrystal substrates with a misorientation angle of $24^\circ$ using
pulsed laser deposition. The as-prepared films showed a transition temperature
$T_c$ between 86 and 87\,K. A 50\,nm thick gold layer providing for contact
pads was evaporated through a shadow mask.  Bridges in a four probe geometry
across the grain boundary with different widths were patterned using a
multi-step optical and electron beam lithography process and ion beam etching.
During the ion beam etching process the sample was cooled using LN$_2$ to
reduce degradation of the film. In this way junctions with widths down to
$\sim 200$\,nm could be fabricated (see Fig.~1). After patterning we always
observed a significant reduction of $T_c$ and an increase of the junction
resistance, which most likely is due to oxygen loss in the thin YBCO films,
especially close to the grain boundary.  This degradation was found to be
caused by the baking of the electron beam resist (PMMA) at $160^\circ$C for
10\,min and could be largely cancelled ($T_c>81$\,K) by an annealing process
in 100\,kPa pure oxygen for 1\,h at $520^\circ$C.

\begin{figure}
\centering{%
\includegraphics[width=8.5cm]{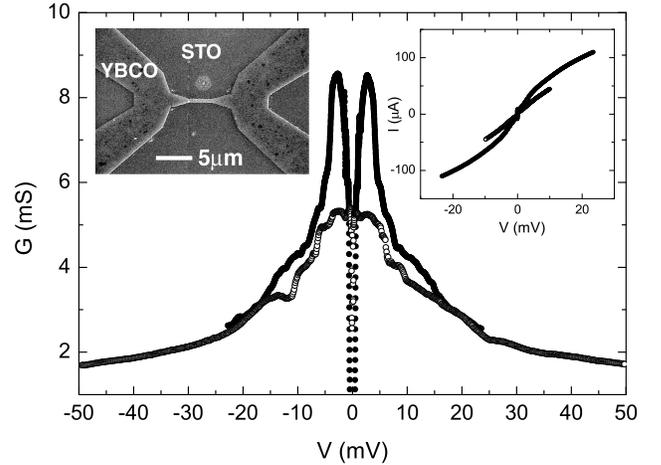}}
\caption{Differential conductance vs voltage of a $30\times600$\,nm$^2$
  bicrystal GBJ at $T=6$\,K. Right inset: Current-voltage curves.  Left inset:
  scanning electron micrograph of a 600\,nm wide GBJ.}
\end{figure}

At 4.2\,K junctions with widths down to 400\,nm showed superconductivity with
critical current densities of several $10^5$A/cm$^{2}$. The inset of Fig.~1
shows typical current-voltage characteristics (IVCs) of a
$30\times600$\,nm$^2$ junction. In the following we present two sets of
transport and noise measurement of this GBJ referred to as {\it 1st run} and
{\it 2nd run}.  The full symbols represent the data obtained in the 1st run
performed immediately after sample preparation. The open symbols show the
result of the 2nd run performed after the sample has been stored in vacuum for
a few days at room temperature. In both runs the junction clearly shows a
hysteretic IVC with critical currents $I_{\rm c1}=8.8\,\mu$A and $I_{\rm c2}=
3.8\,\mu$A, respectively. The differential conductance $G$ measured using
standard lock-in technique is shown in Fig.~1. Both $G(V)$ curves reveal a
strong increase of $G$ with decreasing $V$ (the dip close to zero voltage is
caused by the hysteretic IVC). This is in contrast to a decrease of $G(V)$
expected for tunnel junctions below the gap voltage\cite{Gross97}.  After the
vacuum storage both $I_c$ and $G$ at $V\lesssim 15$\,mV are noticeably reduced
compared to the first run. This most likely is caused by oxygen loss close to
the grain boundary during vacuum storage. For $V\gtrsim 15$\,mV the $G(V)$
curves are nearly identical. The fine structures in the $G(V)$ spectra are
reproducible features which have been observed for different samples up to 20
to 30\,K, where they are continuously smeared out.  Both the increase of $G$
below the gap voltage and the fine structure in the $G(V)$ curves can be
qualitatively understood in terms of the model of Blonder, Tinkham, and
Klapwijk (BTK), assuming a grain boundary barrier with high
transmissivity\cite{Blonder82,Klapwijk82,Loefwander99}. We note that GBJs with
larger misorientation angles have barriers with smaller transmissivity
resulting in tunneling like $G(V)$ curves\cite{Gross97,Alff98a}.

\begin{figure}
\centering{%
\includegraphics[width=8.5cm]{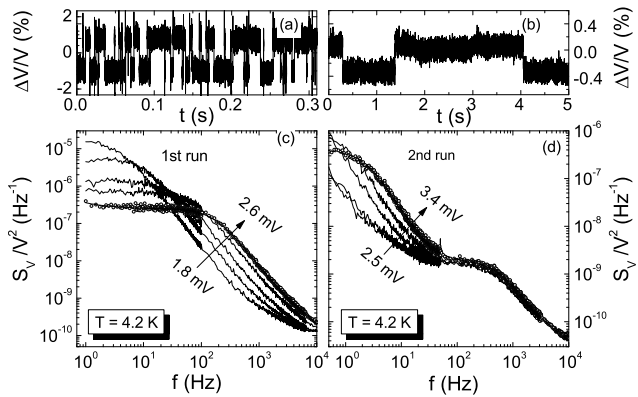}}
\caption{Voltage time traces ((a) and (b)) and noise spectra ((c) and (d))
  before ((a) and (c)) and after ((b) and (d)) vacuum storage of the GBJ.
  Thick solid lines: Fit of Eq.~(\ref{lorentz}) to one of the spectra.}
\end{figure}

The voltage fluctuations were measured with a low-$T_c$ dc SQUID
amplifier\cite{Marx95a}.  In Fig.~2 (c) and (d) two series of voltage noise
spectra are shown which have been measured at 4.2\,K for different junction
voltages. Fig.~2 (a) and (b) display examples of time traces $\Delta V(t)$
recorded at the same temperature for $V=2.5$\,mV clearly showing distinct
random telegraph switching (RTS) signals with relative switching amplitudes
$\Delta V/V$ up to 3\% in the 1st run and approximately 0.5\% in the 2nd run.
Assuming a homogeneous current density and a total suppression of the local
conductance around the responsible noise centers these switching amplitudes
correspond to an impact radius of 7.6\,nm and $1-3$\,nm, respectively. The
two-level signals dominated $\Delta V(t)$ of both runs up to $T\sim 30$\,K.
However, the extremely large fluctuation observed in the 1st run could not be
detected in the 2nd run. Together with a surface degradation of the sample we
expect a loss or redistribution of oxygen at the grain boundary, leading to
either a direct creation/destruction of noise centers or a current
redistribution around them, to be the most likely explanation for this change
of the noise properties between the two measurements.

Within the accessible frequency range (0.5 to 2$\times$10$^4$\,Hz) all spectra
could be well fitted with a superposition of a few ($n\leq3$) independent
Lorentzian components\cite{Machlup54}
\begin{equation}
S_v(f) = \frac{a}{f}+\sum_{i=1}^{n}\frac{4\tau_{\rm eff}^{(i)}\langle(\delta
V_i)^2\rangle}{1+(2\pi\tau_{\rm eff}^{(i)} f)^2}\,,
\label{lorentz}
\end{equation}
together with a weak $1/f$-background ($a<10^{-13}$\,V$^2$). Each Lorentzian
component represents the contribution of a single two-level fluctuator (TLF)
with a switching amplitude $\Delta V_i$ and mean lifetimes $\tau_u$ and
$\tau_l$ in the upper and lower resistance state, respectively. From the fits
to the spectra we obtained the effective lifetimes
\begin{equation}
\tau_{\rm eff} = 1/(\tau_u^{-1}+\tau_l^{-1})
\label{tau_eff}
\end{equation}
as well as the mean squared fluctuation amplitudes
\begin{equation}
\langle(\delta V)^2\rangle = \left(\frac{\tau_u}{\tau_l}+\frac{\tau_l}{\tau_u}
+2\right)^{-1}(\Delta V)^2
\label{dv}
\end{equation}
of the underlying RTS signals.

\begin{figure}
\centering{%
\includegraphics[width=7cm]{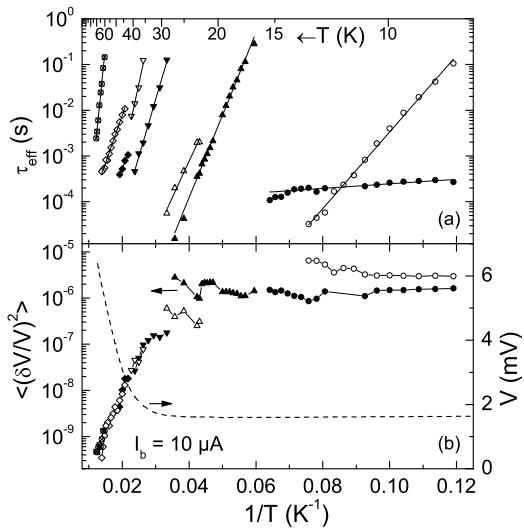}}
\caption{Effective lifetimes $\tau_{\rm eff}$ (a) and mean squared voltage
  fluctuations $\langle(\delta V)^2\rangle$ (b) of a series of TLFs over a
  broad temperature range obtained from fits of Eq.\ (\ref{lorentz}) to the
  noise spectra (2nd run). Identical symbols in both graphs correspond to the
  same fluctuator. Dashed line: junction voltage for $I_b=10\,\mu$A.}
\end{figure}

Fig.~3 shows the temperature dependence of $\tau_{\rm eff}$ and
$\langle(\delta V/V)^2\rangle$ over a broad range of temperatures in an
Arrhenius plot for a series of TLFs of a junction after the vacuum storage.
The measurement was performed at a constant bias current $I_b$ of 10\,$\mu$A.
All fluctuators show a thermally activated behavior in the temperature range
from 8 to 80\,K as has already been observed by Kemen {\it et
  al.}\cite{Kemen99}. By fitting an Arrhenius law $\tau_{\rm eff}=\tau_0
\exp(E_a/k_bT)$ to the data we found activation energies $E_a=2$ to 145\,meV
and attempt times $\tau_0=10^{-11}$ to $10^{-6}$\,s typical for charge
trapping centers in oxide materials\cite{Rogers85}.  Applying a small magnetic
field had no effect on the effective lifetimes and only minor effects on the
mean squared switching amplitude even for $I_b\simeq I_c$. We therefore
exclude magnetic flux jumps in the junction or adjacent film regions as the
origin of the RTS signals as has been proposed for RTS noise in high $T_c$
thin films and SQUIDs\cite{Jung91}.

The temperature dependence of $\langle(\delta V/V)^2\rangle$ as shown in
Fig.~3 shows two striking features. Firstly, there is a crossover from an
almost constant value below $\sim 30$\,K to an exponential decrease for
$T>30$\,K. Secondly, the relative mean squared switching amplitudes
$\langle(\delta V/V)^2\rangle$ of all fluctuators follow a common
temperature dependence in spite of the broad scatter in the effective
lifetimes without any systematic variation with temperature. This indicates
that the population \emph{dynamics} of a trap and its influence on the
charge transport seem to be decoupled. The crossover temperature roughly
corresponds to the temperature where the voltage at $I_b=10\,\mu$A starts
to increase (dashed line in Fig.~3b). Since $\langle(\delta V/V)^2\rangle$
depends on both the relative switching amplitude $\Delta V/V$ and the
ratio $\tau_u/\tau_l$ of the switching times (see Eq.~(\ref{dv})), it is
hard to tell the reason for the exponential decrease in Fig.~3b because
$\tau_u/\tau_l$ exponentially depends on both $T$ and $V$ (cf.\ Fig.~4).
On the other hand, judging from the diverse temperature dependencies of
the individual fluctuators in Fig.~3a and the fact that there were no more
RTS signals resolvable over the apparently featureless noise background in
the time traces at high temperatures, it is reasonable to conclude that
the decrease of $\langle(\delta V/V)^2\rangle$ is to a large extent due to
a decrease of the relative switching amplitude. It is further interesting
to note that this decrease of $\langle(\delta V/V)^2\rangle$ would
coincide with the decrease of the enhanced conductance in the low voltage
regime which may be interpreted in a way that the noise centers in the
junction mainly affect a distinct kind of conductance channels which are
only active at low temperatures. Clearly, a detailed investigation of the
$T$ and $V$ dependence of $\tau_u$ \emph{and} $\tau_l$ is needed to
clarify this issue.

The large RTS signal ($\delta V/V\simeq 3\%$) which occurred during the first
run allowed for a direct analysis of the voltage time traces. Thus the mean
lifetimes $\tau_u$ and $\tau_l$ of both voltage states and the switching
amplitude $\Delta V$ could be determined independently (this is not possible
by analyzing the noise spectra without further assumptions, see
Eqs.~(\ref{tau_eff}) and (\ref{dv})). As the switching times in the two
conductance states were found to be exponentially distributed at specific
values of the voltage and temperature, the mean lifetimes could be determined
by fitting an exponential decay function to the switching time histogram of
the respective state.

\begin{figure}
\centering{%
\includegraphics[width=7cm]{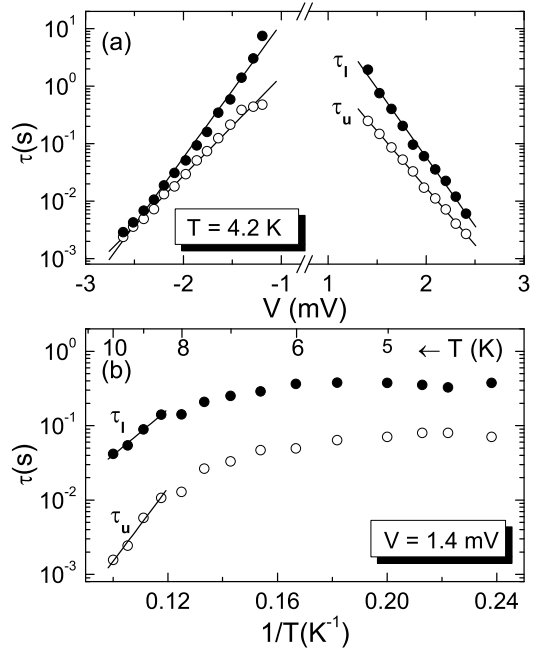}}
\caption{Voltage and temperature dependence of the mean lifetimes in the upper
  ($\tau_u$) and lower voltage state ($\tau_u$), obtained from a direct
  analysis of the time traces of the 1st measurement run.}
\end{figure}

Fig.~4 shows the voltage and temperature dependence of the mean lifetimes in
the upper and lower voltage state at $4.2$\,K and $1.4$\,mV. Both lifetimes
decrease exponentially with the applied voltage at a similar rate. This
confirms our previously reported results for the {\it effective} lifetimes
which have been derived from Lorentzian fits to a series of noise spectra
measured on a 1\,$\mu$m wide $24^\circ$ YBCO-GBJ\cite{Kemen99}. In addition to
that our data show that the \emph{ratio} of the switching times varies
only weakly with the applied voltage and that they are almost independent of
the bias direction. In terms of the trap model this indicates a symmetric
position of the observed trap within the barrier region\cite{Rogers87a}.

From Fig.~4b it becomes obvious that not only the effective mean lifetimes
show a thermally activated behavior at $T\geq 8$\,K but that the lifetimes of
{\it both} voltage states are thermally activated with different kinetic
parameters both lying in the same range found for the effective lifetimes (see
above). Furthermore, for $T\leq 8$\,K both switching times turn over to a $T$
independent behavior indicating a tunneling-like switching in this temperature
range. These observations are in good agreement with our former results
concerning the effective lifetimes at low temperatures\cite{Kemen99}.

In summary, we have fabricated submicron YBCO-GBJs with areas down to
$0.02\,\mu{\rm m}^2$ using electron beam lithography and measured their
electrical transport and noise properties.  The noise spectra and the voltage
vs time traces were found to be dominated by random telegraph switching
signals. The switching kinetics were found to be tunneling-like at $T\lesssim
8$\,K and thermally activated above this value up to $T_c$.  At low $T$ the
mean lifetimes decrease exponentially with the junction voltage. These results
confirm the assumption that the low frequency noise in high-$T_c$ GBJs is
caused by the stochastic capture and release of charge carriers at trapping
centers within the barrier region. In addition, we found a similar mean
squared switching amplitude for all two-level fluctuators in a sample which
follows a common temperature dependence decaying exponentially above about
30\,K.

This work is supported by the Deutsche Forschungsgemeinschaft (Ma 1953/1-1).

\end{document}